\documentclass[twocolumn]{aastex62}
\newcommand{\psr}{PSR~\ensuremath{{\rm J}0636+5128}}

\newcommand{\expnt}[2]{\ensuremath{#1 \times 10^{#2}}}   
\newcommand{\gcc}{\ensuremath{{\mathrm g}\,{\mathrm c\mathrm m}^{-3}}}
\newcommand{\gfilt}{\ensuremath{g^\prime}}
\newcommand{\rfilt}{\ensuremath{r^\prime}}
\newcommand{\ifilt}{\ensuremath{i}}
\newcommand{\tirr}{\ensuremath{T_{\rm irr}}}
\newcommand{\tday}{\ensuremath{T_{\rm day}}}
\newcommand{\tnight}{\ensuremath{T_{\rm night}}}
      
\shorttitle{A Companion to \psr}
\shortauthors{Kaplan et al.}
\accepted{20 July, 2018}
\submitjournal{ApJ}
  
\begin{document}

\title{A Dense Companion to the Short-Period Millisecond Pulsar Binary \psr}
\author[0000-0001-6295-2881]{D.~L.\ Kaplan}
\affiliation{Center for Gravitation, Cosmology and Astrophysics, Department of Physics, University of Wisconsin--Milwaukee, P.O.\ Box 413, Milwaukee, WI 53201, USA}

\author[0000-0002-7261-594X]{K.\ Stovall}
\affiliation{National Radio Astronomy Observatory, 1003 Lopezville
  Road, Socorro, NM 87801, USA}

\author[0000-0002-5830-8505]{M.~H.\ van Kerkwijk}
\affiliation{Department of Astronomy \& Astrophysics, University of Toronto, 50 Saint George Street, Toronto, ON M5S 3H4, Canada}

\author[0000-0002-4223-103X]{C.\ Fremling}
\affiliation{The Oskar Klein Centre, Department of Astronomy,
  Stockholm University, AlbaNova, 10691, Stockholm, Sweden}
\affiliation{Cahill
  Center for Astrophysics, California Institute of Technology,
  Pasadena, CA, 91125, USA}

\author[0000-0002-8811-8171]{A.~G.\ Istrate}
\affiliation{Center for Gravitation, Cosmology and Astrophysics, Department of Physics, University of Wisconsin--Milwaukee, P.O.\ Box 413, Milwaukee, WI 53201, USA}

\correspondingauthor{D.~L.~Kaplan}
\email{kaplan@uwm.edu}

\begin{abstract}
\psr\ is a millisecond pulsar in one of the most compact pulsar
binaries known, with a 96\,min orbital period.  The pulsar mass function
suggests a very low-mass companion, similar to that  seen in
so-called ``black widow'' binaries.  Unlike in most of those, however,
no radio eclipses by material driven off from the companion were seen leading to the possibility that the companion was a degenerate remnant of a carbon-oxygen white dwarf.
We report the discovery of the optical counterpart of its companion in images taken with the Gemini North and Keck~I telescopes.  The companion varies between $\rfilt=25$ and $\rfilt=23$ on the 96\,min orbital period of the binary, caused by irradiation from the pulsar's energetic wind.  We modeled the multi-color lightcurve using parallax constraints from pulsar timing and determine a companion mass of $\expnt{(1.71\pm0.23)}{-2}\,M_\odot$, a radius of $\expnt{(7.6\pm1.4)}{-2}\,R_\odot$, and a mean density of $54\pm26\,\gcc$, all for an assumed neutron star mass of $1.4\,M_\odot$.   This makes the companion to \psr\ one of the densest of the ``black widow" systems.  Modeling suggests that the composition is not predominantly hydrogen, perhaps due to  an origin in an ultra-compact X-ray binary.
\end{abstract}


\section{Introduction}
\object[PSR J0636+5128]{\psr}\ (also called PSR~J0636+5129) was discovered as part of the Green Bank North Celestial Cap
(GBNCC) pulsar survey \citep{2014ApJ...791...67S}.  It was
particularly notable for its short orbital period of only $P_B=96$\,min:
only \object[PSR J1311-3430]{PSR~J1311$-$3430} \citep{2012Sci...338.1314P} and
\object[PSR~J0024-7203R]{PSR~J0024$-$7203R} \citep{2017MNRAS.471..857F} have shorter orbits, and only by
$2.2\,$min and 36\,s, respectively.  It also has a rather low-mass
companion, with a minimum mass of about $7.4\,M_J$
($\expnt{7}{-3}\,M_\odot$, assuming a  pulsar mass of $M_p=1.4\,M_\odot$).
This puts it into the range of ``black widow'' systems
\citep{1988Natur.333..237F,1988Natur.334..686F,1988Natur.334..504K,2013IAUS..291..127R}, where a very low-mass ($\lesssim 10^{-2}\,M_\odot$) companion is in a tight orbit with an energetic
pulsar.  Typically eclipses are seen where ionized material driven off
the companion delays and blocks the radio pulses, although this can be
dependent on inclination.  Systems also often have variations in their
timing parameters suggestive of orbital interactions \citep[e.g.,][]{1994ApJ...426L..85A,1994ApJ...436..312A,1998ApJ...499L.183S,2016MNRAS.462.1029S,2015ApJ...807...18P}.  Initially,
no eclipses or timing variations were seen from
\psr\ \citep{2014ApJ...791...67S} leading to the suggestion that it was instead an inert,
degenerate companion similar to that inferred in the \object[PSR J1719-1438]{PSR~J1719$-$1438}
system \citep{2011Sci...333.1717B}, which has a mass of $\sim 1\,M_J$
and a minimum mean density of $23\,\gcc$.  However, some
black widows show no eclipses or other timing variations
\citep[e.g.,][]{2011AIPC.1357...40H,2015ApJ...813L...4B,2018ApJS..235...37A}, so further investigation of the nature of the  companion to \psr\ was necessary.  

Such distinctions matter because the question of density is used as a
proxy for composition, which is itself used to understand the
formation mechanism for black widow and similar systems.  The
canonical model is that they evolve from low-mass X-ray binary (LMXB)
systems that move to tight orbits and lose considerable mass through
accretion, ejection, and ablation
\citep[e.g.,][]{1992A&A...265...65E,2002ApJ...565.1107P,2012ApJ...753L..33B,2013ApJ...775...27C}.
In contrast, some systems may evolve from ultracompact X-ray binaries
(UCXBs) consisting of a neutron star accreting from a degenerate white
dwarf donor in a compact (orbital period of order an hour) binary
\citep{2003ApJ...598.1217D,2012A&A...537A.104V,2012A&A...541A..22V,2017MNRAS.470L...6S}.
The binary companion would then have primarily a carbon/oxygen (if
originally more massive) or helium composition, compared with hydrogen
in the LMXB scenario, and these can be distinguished at some level
through estimates of density
\citep[e.g.,][]{2014ApJ...791L...5T,2018MNRAS.475..469S}.
A
lower limit for density is possible for these systems by the orbital
period-mean density relation \citep[e.g.,][]{2002apa..book.....F} constraining the density of the Roche lobe.  If
a companion can be identified then further constraints are possible
through estimates of the Roche-lobe filling fraction or companion
radius \citep[e.g.,][]{2014ApJ...791L...5T,2018MNRAS.475..469S}.  In
the case of PSR~J1719$-$1438 such an identification was hampered by
its distance (dispersion measure distance of 1.2\,kpc) and the crowded
field at low Galactic latitude.  However, for \psr\ the higher Galactic latitude makes the search more promising.

Since its discovery, \psr\ has been timed regularly (on a roughly
monthly basis) as part of the North American Nanohertz Observatory for
Gravitational Waves \citep[NANOGrav;][]{2018ApJS..235...37A}.  This
data-set includes a detection of a marginal timing parallax
$\varpi=0.88\pm0.30\,$mas (note that this implies a considerably
larger distance than that determined in \citealt{2014ApJ...791...67S},
and is not consistent with the previous value\footnote{The explanation
  for the difference is likely the longer, higher-quality data-span as
  well as better separation of secular trends in the dispersion
  measure from periodic (parallax) trends.}), and a statistically
significant orbital period derivative $\dot
P_B=\expnt{(2.5\pm0.3)}{-12}$ ($P_B/\dot P_B \approx 77\,$Myr).  As in
other black widow systems this orbital period derivative is unlikely
to come from gravitational radiation, as it is of the wrong sign and
two orders of magnitude too large: general relativity predicts
  $\dot P_B^{\rm GR}=\expnt{-4.3}{-14}$ assuming the nominal masses
  found below \citep[e.g.,][]{2012hpa..book.....L}.  Instead it
likely reflects some mass-loss or other orbital interaction in the
system.  In this paper we report on the identification of the optical
counterpart from observations with Gemini North and Keck~I.  Moreover,
we measure significant orbital modulation coming from irradiation by
the pulsar and use this to estimate the inclination and radius of the
companion.  Note: after submission of this manuscript, we
  became aware of \citet{2018arXiv180704249D} who combine our archival
  Gemini data with their own near-infrared imaging to study the
  companion of \psr.  Our analysis of the lightcurve is broadly
  consistent with that of \citet{2018arXiv180704249D}.

\section{Observations and Reduction}
We observed \psr\ with the Gemini Multi-Object Spectrograph (GMOS;
\citealt{2004PASP..116..425H}) on the 8.1-m Frederick C.\ Gillett
Gemini North telescope on Mauna Kea in Hawaii.  Observations consisted
of $10\times 420\,$s with the \gfilt\ filter and $9\times 420\,$s with the
\rfilt\ filter on the night of 2014~December~21, spanning 69\,min
(0.7\,orbits) in \gfilt\ and 61\,min (0.6\,orbits) in \rfilt.  The
detector was binned $2\times 2$ for a plate scale of  $0\farcs15\,{\rm
  pixel}^{-1}$.  Data were reduced using the GMOS pipeline
\citep{gmos}.  The airmass ranged from 1.2--1.3 (\gfilt) and 1.3--1.5
(\rfilt), while the seeing was about $1\farcs0$ (\gfilt) and
$0\farcs9$ (\rfilt).

We also observed \psr\ with the Low-Resolution Imaging Spectrometer
(LRIS; \citealt{1995PASP..107..375O,2010SPIE.7735E..0RR}) on the 10-m Keck~I telescope 
on Mauna Kea in Hawaii.    Observations consisted of $13\times 300\,$s
using the red side\footnote{Observations using the blue
side with the \gfilt\ filter were processed, but the signal-to-noise was very low (best detections had signal-to-noise of 3.5, compared to 17 for GMOS) and the results were consistent with the GMOS data.  Therefore we use the GMOS \gfilt-band data exclusively.}  of the instrument and the \ifilt\ filter on the night of
2018~January~18, spanning 87\,min (0.9\,orbits). The airmass ranged
from 1.2--1.4, and the seeing was about $1\farcs9$.  Data were reduced
using the \texttt{LPIPE} package.

Once the raw images were reduced, we made sure that they were
registered astrometrically by comparing with stars from the Pan STARRS
$3\pi$ survey \citep[PS1;][]{cmm+16}.  We easily identified a variable
optical source at the radio timing position of the pulsar
(Fig.~\ref{fig:image}). We measured fluxes (rejecting
cosmic rays with a simple threshold) in a single
aperture with a constant radius for each instrument/filter combination that was close to the seeing FWHM on 11 selected stars
from PS1 and also the timing position of the pulsar.  These stars were
selected to be bright, well-isolated, and not saturated.  All of the
reference stars were used to determine relative zero-points within
each filter, and absolute photometry was referenced to the star
PSO~J063600.942+512838.878, chosen because it was bright, nearby, and not in a crowded region; attempts to use an ensemble for more accurate absolute photometry had difficulties regardless of whether the photometric standard was PS1, the Sloan Digital Sky Survey \citep{2018ApJS..235...42A}, or others.  Based on observations of the ensemble of
reference stars we believe the relative photometry to be accurate to
about $0.02$\,mag (\gfilt\ and \rfilt) or 0.01\,mag (\ifilt).  The
absolute photometry is likely to only be accurate to about 0.1\,mag,
since we referenced it to a single star and do not include color terms
relating to different photometric systems.

Finally, we corrected the image times to the solar system barycenter
using routines in \texttt{astropy} and computed the orbital phase of
the midpoint of each image.

\begin{figure}
  \plotone{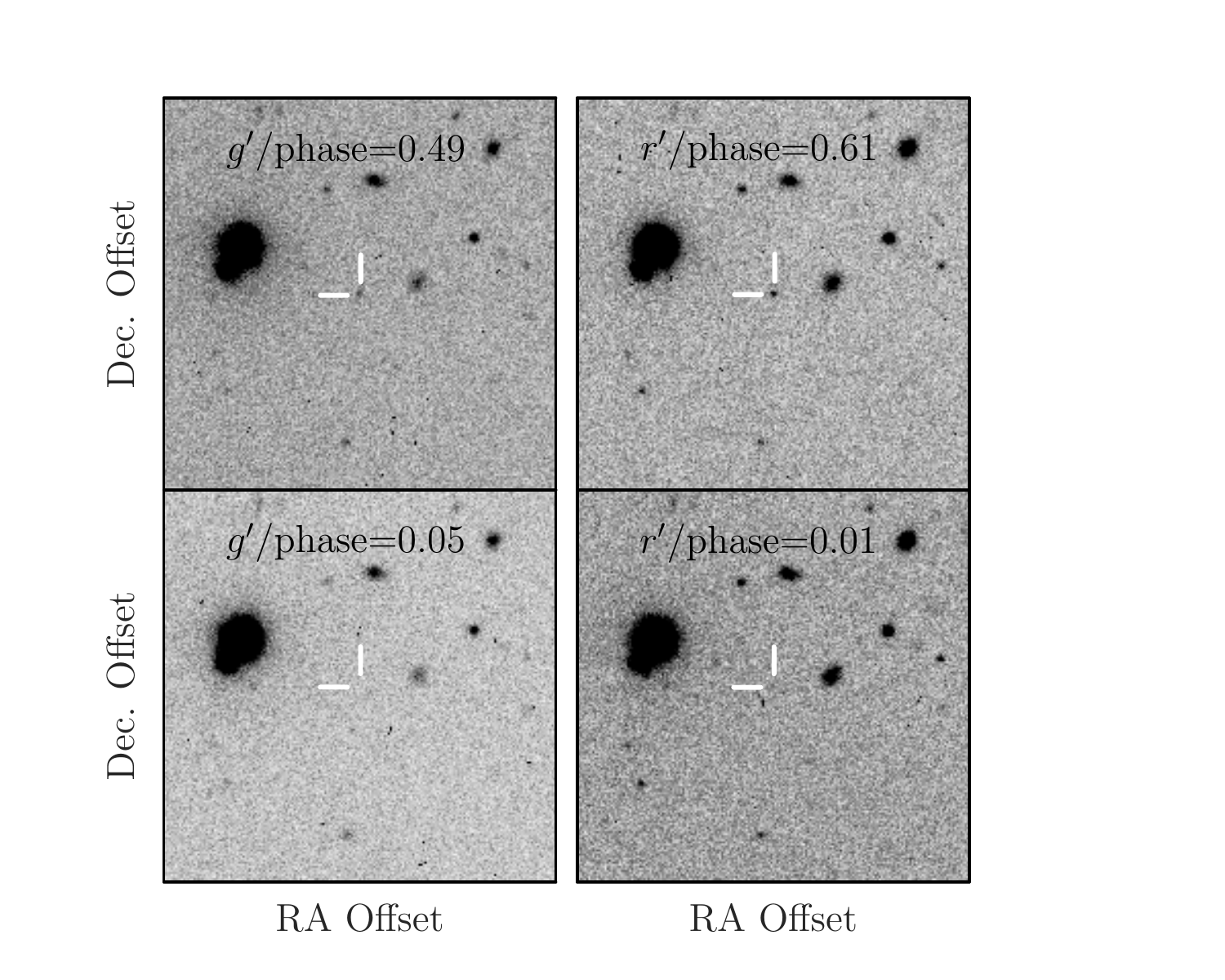}
  \caption{Gemini-N GMOS images of \psr.  We show \gfilt-band (left)
    and \rfilt-band (right) for two different orbital phases, close to
    photometric maximum/phase of 0.5 (top) and close to photometric
    minimum/phase of 0 (bottom).   The images are $60\arcsec$ on a
    side, with north up and east to the left.  The tick marks indicate
    the position of \psr.}
  \label{fig:image}
\end{figure}

\section{Light Curve Fitting}
We modeled the \gfilt-, \rfilt-, and \ifilt-band lightcurves of \psr\ using
\texttt{Icarus} \citep{2012ApJ...748..115B}.  The model consisted of a
binary with possible irradiation, ellipsoidal modulation, and Doppler
boosting.  We assumed corotation of the companion, and took the
gravity darkening coefficient to be 0.08 (appropriate for convective
envelopes, which is likely the case in the effective temperature range
that we found).
The free parameters are the inclination $i$, Roche-lobe
filling factor $f$, backside temperature \tnight, irradiated temperature
excess $\tirr$, extinction $A_V$, and parallax $\varpi$; we held
the neutron star mass to be fixed during the fitting.  The irradiated
temperature \tirr\ is related to the front-side temperature \tday\ (facing
the pulsar) and the backside temperature \tnight\ by
$\tday^4=\tnight^4+\tirr^4$.  For the inclination we used a prior
distribution that was flat in $\cos i$; the priors for $f$, \tnight,
and \tirr\ were uniform over $[0,1]$, $[0,20000]$\,K, and $[0,20000]$\,K,
respectively.  We used a prior distribution for $A_V$ informed by the
three-dimensional Galactic extinction model of
\citet{2018arXiv180103555G}, which gives $A_V=0.25\pm0.06$\,mag at the nominal
distance of 1.2\,kpc.  For the parallax we used a normal prior determined by
the radio timing observations, ${\cal N}(0.88\,{\rm mas},0.30\,{\rm
  mas})$.  Finally, we added an additional prior with $p(\varpi) \propto
\varpi^{-4}$ to account for Lutz-Kelker bias \citep{lk73,vlm10} in this
low-significance measurement.  e note that underlying this prior is an
assumption of a constant space density for binaries like \psr, which is
unlikely to hold.  As for other types of MSP binaries, more realistic
spatial distributions would likely lead one to one infers somewhat smaller
distances \citep[e.g.,][]{vwc12,2016A&A...591A.123I,2018arXiv180606076J}.
We  allowed for an additional systematic offset that was free for each photometric band with an uncertainty of 0.1\,mag, as discussed above.  The $\chi^2$ from this band offset was added to the $\chi^2$ for the individual photometric points.

With this model we performed a Markov Chain Monte Carlo (MCMC) fit
using the  affine invariant MCMC ensemble sampler \texttt{emcee}
\citep{2013PASP..125..306F}.  We started {400} ``walkers'' in
the 6-dimensional parameter space and allowed them to evolve for
{100} steps to achieve ``burn-in.''  We then reset the sampler
and evolved for a further {1000} steps, saving all of the samples for a total of 40,000 MCMC samples.

In Table~\ref{tab:results} we give the best-fit values of the
parameters for three different assumed neutron star masses, taken to be the medians of the resulting posterior
probability distribution functions.  We give both the actual fitted
parameters and derived parameters: the companion mass $M_c$ and radius
$R_c$, the companion mean density $\rho_c$, and the irradiation
efficiency $\eta$ defined by:
\[
\sigma \tirr^4 = \eta \frac{\dot E}{4\pi a^2}
\]
where $\dot E$ is the spin-down luminosity of the pulsar, $a$ is
the inferred orbital separation, and $\sigma$ is the Stefan-Boltzmann constant. 

Overall we were able to achieve a reasonable fit, and we show the best-fit lightcurve for $1.4\,M_\odot$ in
Figure~\ref{fig:lc}.  The fit yields $\chi^2=48.4$ for
23 degrees-of-freedom including systematic offset terms for each photometric band of 
 $0.1\,$mag each.  We give the
best-fit values and  uncertainties
(determined from the inner-quartile range, which is more robust to
outliers than other methods) in Table~\ref{tab:results}. We have increased the uncertainties in the fitted parameters from Table~\ref{tab:results} by the square root of the reduced $\chi^2$ to account for any underestimated uncertainties or modeling errors.  The band offsets were small, consistent with our estimates of the systematic uncertainties.  The dereddened  color varies from $\gfilt-\ifilt=1.45$ at photometric maximum to $\gfilt-\ifilt=1.95$ at photometric minimum (assuming $A_V=0.25\,$mag), which implies changing from spectral type K4 to M0 \citep{2007AJ....134.2398C} or effective temperatures ranging from 4600\,K to 3800\,K.  We find no evidence for dramatic flares or other stochastic variations such as those seen in PSR~J1311$-$3430 \citep{2012ApJ...760L..36R}.
The photometric variability is dominated by irradiation of the
companion by the pulsar (fractional amplitude of about 70\%), which has a period equal to the binary
period.  We see no evidence of ellipsoidal modulation (at twice the
orbital period), which is consistent with the modest Roche-lobe
filling of the companion.  The ellipsoidal modulation should be at
most about 2\% (based on \citealt{2012ApJ...748..115B}), consistent
with the amplitudes of fitted sinusoids.  A final potential cause of
periodic modulation is Doppler boosting
\citep{2000MNRAS.317L..41M,2003ApJ...588L.117L,2007ApJ...670.1326Z}, which is at the
orbital period although at a different phase compared to irradiation, but
despite the high inferred velocity for the companion (inferred radial velocity
amplitude of $589\,{\rm km\,s}^{-1}$) this is only expected to be
1.6\% even in the \gfilt-band,  below our detection limits of 18\% (2-$\sigma$).

The full fit results are shown in Figure~\ref{fig:corner}.
The best-fit values of $A_V$ and $\varpi$
agreed with the prior distributions.  There is a small tail of inclinations that extends to high values, leading to a tail in the distributions of companion mass and density, but  only about 10\% of the probability has $i>40\degr$.
None of the other fitted
parameters had significant bimodalities or other issues.  The backside temperature formally extends to low values, even as low as 0\,K.  However, since the inclination is largely face-on we never see just the backside of the companion so the lowest area-average temperature is considerably higher, consistent with the colors above.   Since a
number of black widow systems have been observed to have high neutron
star masses \citep[e.g.,][]{2011ApJ...728...95V,2012ApJ...760L..36R,2014ApJ...793...78S}, our results
are given both for a canonical neutron star mass of $1.4\,M_\odot$ as
well as higher values of $1.8\,M_\odot$ and $2.0\,M_\odot$.  However,
we do not see a significant shift in our fit results for those other
values, as $M_c$ can just scale up along with a small increase in distance to compensate.  In what follows we will primarily discuss the results for
$1.4\,M_\odot$.

\begin{figure}
  \plotone{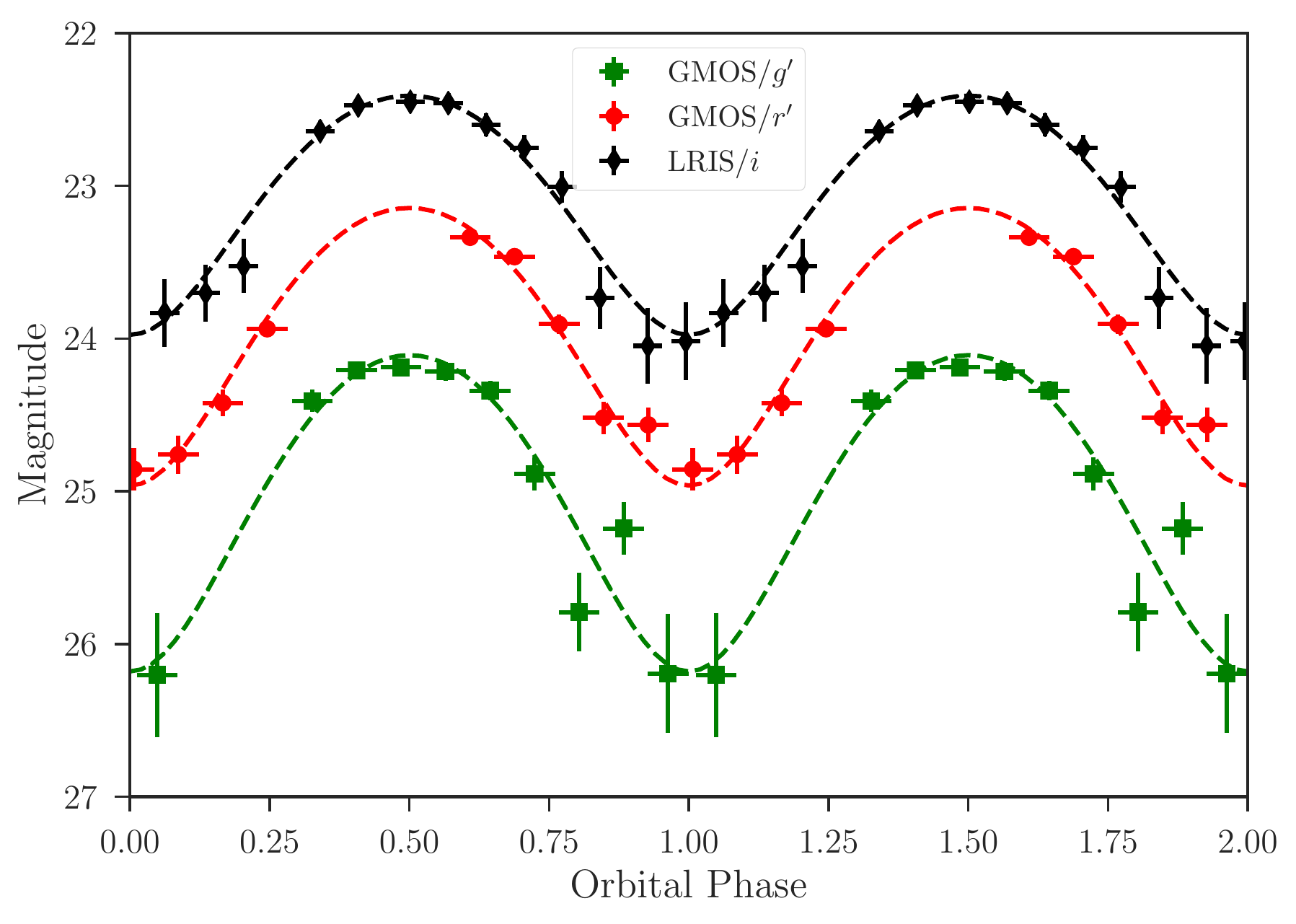}
  \caption{Best-fit \texttt{Icarus} lightcurve model for \psr\ repeated twice for clarity,
    assuming a pulsar mass of $1.4\,M_\odot$.  Note
    that orbital phases of 0 and 0.5 corresponds to conjunction.  We
    show the GMOS \gfilt- and \rfilt-bands (green squares and red
    circles, respectively) and the LRIS \ifilt-band (black diamonds)
    together with the best-fit model lightcurves.  The best-fit parameters are in Table~\ref{tab:results}.}
  \label{fig:lc}
\end{figure}

\begin{figure*}
  \plotone{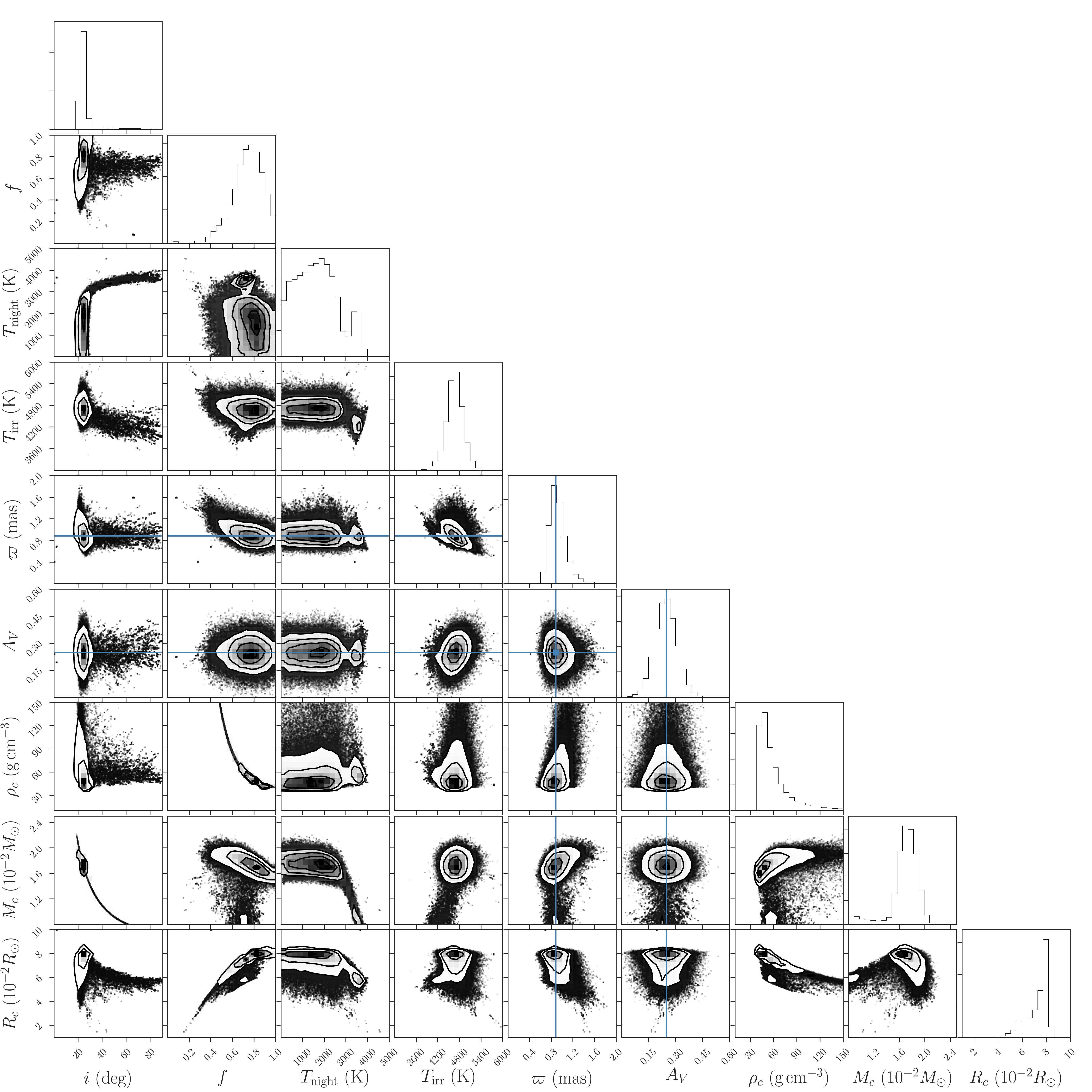}
  \caption{Corner plot showing the MCMC results for fitting a
    lightcurve to the data for \psr, assuming a pulsar mass of
    $1.4\,M_\odot$.  We show the distributions for the fitted parameters: inclination
    $i$, Roche-lobe filling fraction $f$, backside temperature
    \tnight, irradiated temperature \tirr, parallax $\varpi$,
    extinction $A_V$.  We also show distributions for three derived
    parameters: the mean density of the companion $\rho_c$, the mass
    of the companion $M_c$, and the radius of the companion $R_c$.
    For the parallax and extinction the vertical/horizontal lines show
    the means of the prior distributions determined from other sources.}
  \label{fig:corner}
\end{figure*}

\begin{deluxetable}{l c c c}
  \tablecaption{MCMC Lightcurve Fit Results\label{tab:results}}
  \tablehead{
    \colhead{Parameter} & \colhead{$M_p=1.4\,M_\odot$} &
    \colhead{$M_p=1.8\,M_\odot$} & \colhead{$M_p=2.0\,M_\odot$}
  }
  \startdata
  \cutinhead{Fitted Parameters}
$i$ (deg)\tablenotemark{a} \dotfill & $24.3 \pm 3.5$\phn & $24.5 \pm 3.8$\phn & $24.6 \pm 4.2$\phn\\
$f$\tablenotemark{b} \dotfill & $0.75 \pm 0.20$ & $0.75 \pm 0.19$ & $0.74 \pm 0.19$\\
\tnight\ (K)\tablenotemark{c} \dotfill &$1643 \pm 1561$\phn & $1709 \pm 1638$\phn & $1706 \pm 1758$\phn\\
\tirr\ (K)\tablenotemark{c} \dotfill &$4671 \pm 324$\phn & $4668 \pm 330$\phn & $4641 \pm 346$\phn\\
$\varpi$\ (mas)\tablenotemark{d} \dotfill &$0.90 \pm 0.21$ & $0.86 \pm 0.19$ & $0.84 \pm 0.18$\\
$A_V$\ (mag)\tablenotemark{e} \dotfill &$0.25 \pm 0.09$ & $0.25 \pm 0.09$ & $0.24 \pm 0.09$\\
\cutinhead{Derived Parameters}
$\rho_c$\ (g\,cm$^{-3}$)\tablenotemark{f} \dotfill &$54 \pm 26$ & $53 \pm 26$ & $54 \pm 27$\\
$M_c$\ ($10^{-2}M_\odot$)\tablenotemark{f} \dotfill &$1.71 \pm 0.23$ & $1.94 \pm 0.28$ & $2.15 \pm 0.34$\\
$R_c$\ ($10^{-2}R_\odot$)\tablenotemark{f} \dotfill &$7.6 \pm 1.4$ & $7.9 \pm 1.4$ & $8.1 \pm 1.8$\\
$\eta$\tablenotemark{c}\ \dotfill &$0.18 \pm 0.05$ & $0.20 \pm 0.06$ & $0.22 \pm 0.07$\\
\tday\ (K)\tablenotemark{c} \dotfill &$4726 \pm 293$\phn & $4730 \pm 287$\phn & $4708 \pm 283$\phn\\
$d$ (kpc)\tablenotemark{d} \dotfill &$1.11 \pm 0.25$\phn & $1.17 \pm 0.26$\phn & $1.19 \pm 0.26$\phn\\
\enddata
\tablecomments{The values quoted here are the medians of the posterior
  probability distributions plus 1-$\sigma$ confidence limits
  determined from the inner quartile range scaled up by the square root of the reduced $\chi^2$.}
\tablenotetext{a}{System inclination.}
\tablenotetext{b}{Roche-lobe filling factor, defined as the radius of
  the companion facing the pulsar (the ``nose") divided by the distance to the L1
  Lagrange point.}
\tablenotetext{c}{Backside temperature \tnight\ and irradiated
  temperature \tirr.  We also give frontside temperature \tday,
  where $\tday^4=\tnight^4+\tirr^4$ and irradiation efficiency $\eta$
  where $\sigma \tirr^4=\eta\dot E/4\pi a^4$, with $\dot
  E=\expnt{5.5}{33}\,{\rm erg\,s}^{-1}$ the
  spin-down luminosity and $a$ the orbital separation.   $\dot E$ is corrected for the proper motion  \citep{1970SvA....13..562S} and assumes a moment of inertia of $10^{45}\,{\rm g\,cm}^2$.}
\tablenotetext{d}{The parallax $\varpi$ and distance $d$.}
\tablenotetext{e}{The $V$-band extinction.}
\tablenotetext{f}{The companion mass $M_c$, companion radius $R_c$,
  and mean density $\rho_c$ determined from the assumed neutron star
  mass and inclination.}
\end{deluxetable}

\section{Discussion and Conclusions}
The observed orbital period derivative $\dot P_B$, is of the wrong
sign and two orders of magnitude too large to be explained by emission
of gravitational radiation.  Instead we examine whether or not it
could be caused by mass-loss from the system.  First we correct $\dot P_B$ (and $\dot P$) 
 for the \citet{1970SvA....13..562S} effect using our distance estimate, and find that only a 3\% correction is needed.  
Scaling the mass-loss with the orbital period change, 
\[
\dot M_c\sim(M_c+M_p) \frac{\dot P_B}{P_B} 
\]

we find $\dot M_c\sim 10^{-8}\,M_\odot\,{\rm yr}^{-1}$.
  This is a plausible amount
of mass-loss for removing the majority of the mass of the companion in
considerably less than a Hubble time after the end of mass transfer. However, it is four
orders of magnitude larger than the mass-loss rate expected for pulsar
irradiation \citep{1992MNRAS.254P..19S}, although the irradiation efficiency is  very similar to those found by \citet{2013ApJ...769..108B} for
a number of other systems.  Instead $\dot P_B$ could originate
in secular orbit interactions such as those seen in other black widow
systems \citep[e.g.,][]{1994ApJ...426L..85A,1994ApJ...436..312A,1998ApJ...499L.183S}.
Further timing to search for higher-order derivatives would be conclusive.
Unlike in other black widow systems,  radio eclipses
have not been detected from this system \citep{2014ApJ...791...67S}, but that can be understood by the relatively face-on
inclination determined above \citep[also see][]{2018arXiv180704249D}.

As discussed in \citet{2014ApJ...791...67S}, the minimum mean density
inferred for the companion of \psr\ is about $43\,\gcc$, almost a factor of two larger
than that inferred for the companion of PSR~J1719$-$1438.  This estimate assumes Roche
lobe filling: our smaller filling factor implies an even higher companion density  of $\approx 54\,\gcc$.  

Our estimates for the companion's  mass and radius place
it right in the region predicted by \citealt{2003ApJ...598.1217D} for
systems with orbital periods of about 90\,min.  It is slightly smaller
and denser than giant planets in this mass range
\citep[e.g.,][]{2015ApJ...810L..25H} which have densities $\approx 20\,\gcc$, but this could be a combination
of a different composition (more C/O-rich, as suggested by
\citealt{2011Sci...333.1717B} for PSR~J1719$-$1438), other internal
differences, or just measurement error.  We compare with models
generated using Modules for Experiments in Stellar Astrophysics
\citep[MESA;][version 10398]{2011ApJS..192....3P}.  Hydrogen models are 
based on the \texttt{brown\_dwarf test\_suite} case. 
The models with helium composition were created starting from a white
dwarf model of $0.35\,M_\odot$ from  the \texttt{white\_dwarf\_models}
database and relaxing the mass until the desired mass is obtained.
The helium model is shown at an effective temperature
of 3,000\,K (similar to  the upper limit on the backside temperature of \psr), while the
hydrogen model is at 2,000\,K since they typically started at
temperatures cooler than 3,000\,K; a higher temperature would tend to \textit{decrease} the density of the hydrogen models, leading to a worse match.  Both of those model tracks are
similar and largely parallel to a model at a constant radius, which is
to be expected since sources have a constant radius for a wide span 
of mass in this range owing to the transition from degeneracy support to Coulomb support.  At an effective temperature of 3,000\,K
\psr\ appears to have a composition with somewhat higher density than
pure hydrogen (similar to helium), which suggests that it could be the
remnant of a helium white dwarf, perhaps indicating a UCXB origin \citep{2017MNRAS.470L...6S}.

Overall, as shown in
Figure~\ref{fig:density}, \psr\ appears to have one of the highest
mean densities of any black widow system.  However, the majority of densities
are lower limits as they assume the system to be Roche-lobe
filling and a number of  systems (especially those in globular
clusters) do not have direct constraints.  For instance,
PSR~J1544+4937 could have a density as high as 500\,\gcc\ (suggesting
an origin in a UCXB system), although the unconstrained distance means
that it could also be a factor of 20 smaller \citep{2014ApJ...791L...5T}.

In terms of previous evolution, our estimate of the mean
density of $54\,\gcc$ is consistent with the remnant of a
helium-core white dwarf that has been ablated, evaporated,
and/or accreted by the pulsar. It is possible the same holds
for the companion to PSR~J1719$-$1438, i.e., that it has a
helium composition instead of the carbon-rich one favored by  \citet{2011Sci...333.1717B}. If so, it would have a similar radius,
since for these masses radius is predicted to be nearly
independent of mass.  If it also were equally hot, it would
have been seen by  \citet{2011Sci...333.1717B}, since the distances
are similar as well and there is only $\sim\!1\,$mag of
excess extinction.  At first glance that might suggest the
companion of PSR~J1719$−$1438 is in fact smaller and thus made
of denser material.  However, the irradiation in this source
is at least 4 times lower than for \psr, which
would lead to a $\gtrsim30$\% decrease in the irradiated
temperature and therewith to a $\sim\!1.5\,$mag decrease in
maximum companion optical brightness.  This would be
consistent with the non-detection of \citet{2011Sci...333.1717B}.

Unfortunately, \psr\ is somewhat faint for optical spectroscopy which could be used to determine the mass ratio and, in conjunction with modeling such as that presented here, determine the neutron star mass \citep{2011ApJ...728...95V,2012ApJ...760L..36R}.  This is especially true since observations would need to cover only a small fraction of the orbital period in order to not suffer too much orbital smearing: alternate techniques such as ``trailed" spectroscopy \citep[e.g.,][]{2015ApJ...804..115R} are possible, but at two magnitudes fainter than PSR~J1311$-$3430 it will still be difficult.

\begin{figure*}
  \plotone{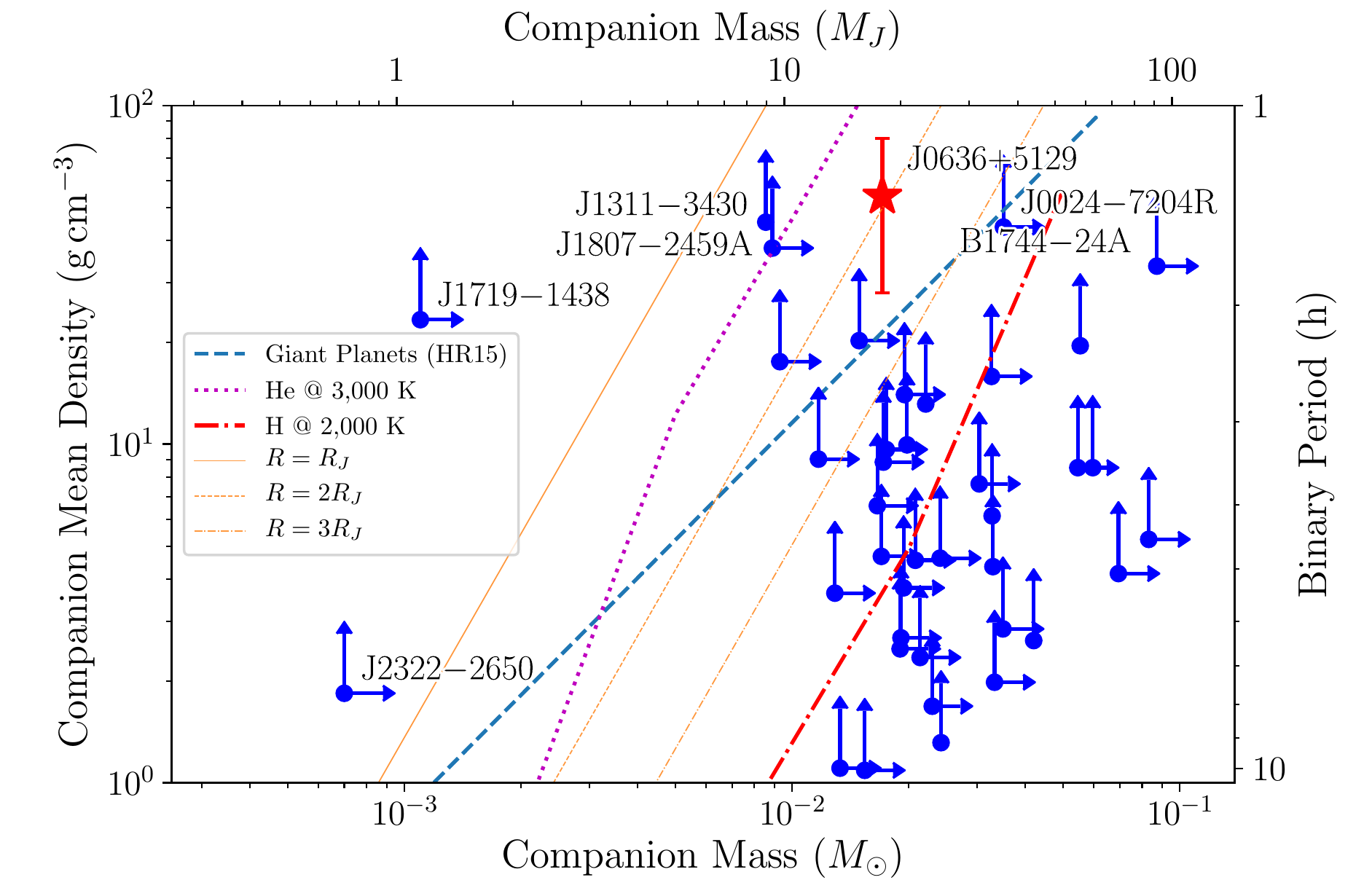}
  \caption{Companion mass versus companion mean density for all
    ``ultra-light'' systems in the ATNF pulsar catalog
    \citep{mhth05,mhtt16}.  For the majority of systems, densities are
    from the orbital period-mean density relation only
    \citep{2002apa..book.....F}, and the companion masses are the
    minimum companion masses assuming a $1.4\,M_\odot$ neutron star.
    For a limited number of systems with published inclination
    constraints  we have updated the companion masses: PSRs
    \object[J0023+0923]{J0023+0923}, \object[PSR J2256-1024]{J2256$-$1024}, \object[PSR J1810+1744]{J1810+1744}, and \object[PSR J2215+5135]{J2215+5135} from
    \citet{2013ApJ...769..108B}, \object[PSR J2129-0429]{PSR~J2129$-$0429} from
    \citet{2016ApJ...816...74B}, \object[PSR~J1301+0833]{PSR J1301+0833} from
    \citet{2016ApJ...833..138R}, \object[PSR B1957+20]{PSR~B1957+20} from
    \citet{2007MNRAS.379.1117R}, \object[PSR J2051-0827]{PSR~J2051$-$0827} from
    \citet{1999ApJ...510L..45S}, \object[PSR J1953+1846A]{PSR~J1953+1846A} from
    \citet{2015ApJ...807...91C},
    \object[PSR J1544+4937]{PSR~J1544+4937} from \citet{2014ApJ...791L...5T}, and PSR~J1311$-$3430 from
    \citet{2015ApJ...804..115R}.  For \psr\ (star) the numbers are
    from this paper.  Select systems are labeled.  The mass
    uncertainties correspond to the range of inclinations from
    $60\degr$ to $90\degr$, while the density uncertainties are a
    factor of 1.5.  We also plot the empirical fit to giant planets
    and brown dwarfs from \citet[][dashed line]{2015ApJ...810L..25H}, models of a
    helium white dwarf remnant at an effective temperature of 3000\,K
    (dotted line), a hydrogen brown dwarf at 2000\,K (dot-dashed
    line).  For reference, we also show source with radii of 1, 2, and 3 times that of Jupiter (thin solid, dashed, and dot-dashed lines).}
  \label{fig:density}
\end{figure*}


\acknowledgements \emph{Acknowledgments.}  Partially based on observations
obtained at the Gemini Observatory under program GN-2014B-Q-81 (PI:
Stovall), which is operated by the Association of Universities for
Research in Astronomy, Inc., under a cooperative agreement with the
NSF on behalf of the Gemini partnership: the National Science
Foundation (United States), the National Research Council (Canada),
CONICYT (Chile), Ministerio de Ciencia, Tecnolog\'{i}a e
Innovaci\'{o}n Productiva (Argentina), and Minist\'{e}rio da
Ci\^{e}ncia, Tecnologia e Inova\c{c}\~{a}o (Brazil). IRAF is
distributed by the National Optical Astronomy Observatory, which is
operated by the Association of Universities for Research in Astronomy
(AURA) under a cooperative agreement with the National Science
Foundation. Some data presented herein were obtained at the W. M. Keck
Observatory, which is operated as a scientific partnership among the
California Institute of Technology, the University of California and
the National Aeronautics and Space Administration. The Observatory was
made possible by the generous financial support of the W. M. Keck
Foundation.  The authors wish to recognize and acknowledge the very
significant cultural role and reverence that the summit of Mauna Kea
has always had within the indigenous Hawaiian community.  We are most
fortunate to have the opportunity to conduct observations from this
mountain.  Support for DLK and KS was provided by the NANOGrav NSF Physics Frontiers
Center award number 1430284. A.G.I. acknowledges support from the NASA Astrophysics Theory Program through NASA grant NNX13AH43G.

\facility{Gemini North (GMOS); Keck:I (LRIS)}

\software{Astropy \citep{2013A&A...558A..33A}, corner \citep{corner}, emcee
  \citep{2013PASP..125..306F} Icarus
  (\url{https://github.com/bretonr/Icarus}), LPIPE
  (\url{http://www.astro.caltech.edu/~dperley/programs/lris/manual.txt}), MESA \citep{2011ApJS..192....3P},
photutils \citep{larry_bradley_2017_1039309}}.


\end{document}